\documentclass[aip,jap,twocolumn,superscriptaddress,a4paper,10pt]{revtex4-1}
\usepackage{amsmath,amssymb,amsfonts}
\usepackage{gensymb} 
\usepackage{epsfig}
\usepackage{graphicx}
\usepackage[dvipdfmx]{hyperref}
\hypersetup{colorlinks=true,citecolor=blue,linkcolor=blue,urlcolor=blue}
\usepackage[all]{hypcap}
\usepackage{color}

\begin{document}
\title{\large{ Optical Absorption and Emission of Silicon Nanocrystals: from Single to Collective Response }}

\author{\firstname{Roberto} \surname{Guerra}}
\affiliation{Centro Interdipartimentale En\&Tech, Universit\`a degli Studi di Modena e Reggio Emilia, via Amendola 2 Pad.\ Morselli - 42122 Reggio Emilia, Italy.}

\author{\firstname{Francesco} \surname{Cigarini}}
\affiliation{Dipartimento di Scienze e Metodi dell'Ingegneria, Universit\`a degli Studi di Modena e Reggio Emilia, via Amendola 2 Pad.\ Morselli - 42122 Reggio Emilia, Italy.}

\author{\firstname{Stefano} \surname{Ossicini}}
\affiliation{Centro Interdipartimentale En\&Tech, Universit\`a degli Studi di Modena e Reggio Emilia, via Amendola 2 Pad.\ Morselli - 42122 Reggio Emilia, Italy.}
\affiliation{Dipartimento di Scienze e Metodi dell'Ingegneria, Universit\`a degli Studi di Modena e Reggio Emilia, via Amendola 2 Pad.\ Morselli - 42122 Reggio Emilia, Italy.}

\begin{abstract}
\textbf{ 
We report on the possibility of describing the absorption and emission characteristics of an ensemble of silicon nanocrystals (NCs) with realistic distributions in the NC size, by the sum of the reponses of the single NCs. The individual NC responses are evaluated by means of \textit{ab initio} theoretical calculations and the summation is performed by taking into account the trend of the optical properties as a function of NC size and oxidation degree. The comparison with experimental results shows a nice matching of the spectra, also without any tuning of the parameters. Finally, the possibility of adapting the model in order to reproduce the experimental data is explored and discussed.
}
\end{abstract}
\maketitle

\section{Introduction}\label{sec.intro}
\noindent Silicon Nanocrystals (Si-NCs) have attracted a lot of interest in the latest years, due to their large applicability. To date Si-NCs have been employed in several fields like nanophotonics,\cite{chen,park,hannah} photovoltaics,\cite{kim_APL95,yuan,timmerman} thermoelectrics,\cite{wang} medical screening,\cite{fujii} and others.
One of the most challenging aspect of Si-NCs concerns the high sensitivity of the measured response to the precise structural configuration of the NC and of its surrounding environment. In fact, size, shape, interface, defects, impurities, embedding medium, and cristallinity level, among others, constitute a set of mutually-dependent parameters that drastically determine the opto-electronic properties of the NCs. Many theoretical and experimental works have contributed to characterize the connection between the above parameters and the observed NC response.
While the theoretical approach is more suitable to deal with single NCs, especially when making use of simulations at the atomistic (\textit{ab-initio}) level, experiments usually make use of samples containing a large number of different NCs, making the identification of the most active configurations a non trivial task. Therefore, despite the tremendous advances of the latest years, a direct comparison between theoretical simulations and experimental observations is still a complicated task.
\\In the present work we try to fill up the gap by extending the theoretical calculations performed on individual NCs to realistic ensembles made by a large number of NCs, in order to provide a connection with the experimental data. Following the superposition principle we aim at evaluating the optical absorption/emission of an ensemble of NCs as the sum of the absorptions/emissions of the individual NCs. The main approximation regards the absence of NC-NC interaction mechanisms, that for H- or OH-terminated NCs, or for embedded Si/SiO$_2$ NCs implies NC-NC distances larger than about 0.5\,nm.\cite{timmerman,govoni,seino} The latter conditions can be satisfied in real embedded or freestanding NC samples by varying the silicon excess or the NC concentration, respectively.
\\It is worth to note that embedded systems present strained bonds at the interface region, of magnitude proportional to the difference between the lattice spacing of Si and that of the embedding medium. Such strain has been recognized as a crucial factor, strongly concurring with others in the determination of the NC properties. In particular, red-shifts of as much as $\sim$1\,eV have been calculated in the absorption spectrum of Si-NCs embedded in a SiO$_2$ matrix w.r.t.\ their freestanding counterparts,\cite{PRB2} while photoluminescence (PL) experiments report red-shifts up to 0.2\,eV.\cite{kusova}
Clearly, an accurate description of an ensemble of embedded NCs cannot ignore the influence of the matrix-induced strain. For this reason the results presented in this work should only indirectly be interpreted in terms of embedded NCs. In principle, a connection of strain with other NC parameters is possible, as already tempted in Ref.~\onlinecite{PRB2}, and could easily be applied to the present method with the purpose of describing ensembles of embedded NCs.

\subsection*{Structures and Methods}\label{sec.struct-methods}
\begin{figure*}[t!]
  \centering
  \includegraphics[draft=false,width=\textwidth,angle=0]{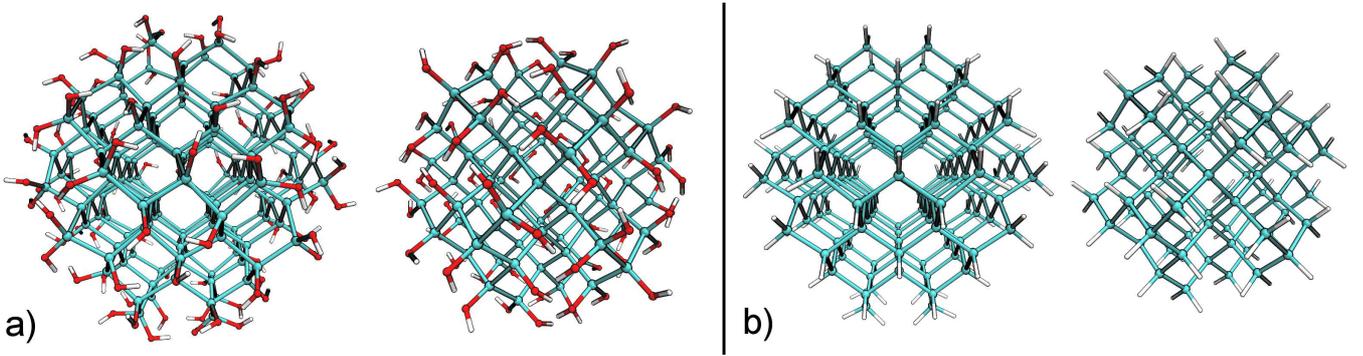}
  \caption{\small Example of OH-terminated (a) and H-terminated (b) NCs after ionic relaxation. From left to right: Si$_{147}$(OH)$_{100}$, Si$_{87}$(OH)$_{76}$, Si$_{147}$H$_{100}$, Si$_{87}$H$_{76}$. Si, O, and H atoms are represented in cyan, red, and white, respectively. }\label{fig.structures}
\end{figure*}
\noindent Structural, electronic and optical properties have been obtained by full \textit{ab-initio} calculations in the framework of density functional theory (DFT) using the {\small ESPRESSO} package.\cite{espresso} Calculations have been performed using norm-conserving pseudopotentials within the local-density approximation (LDA). An energy cutoff of 60 Ry on the plane-wave basis set has been considered. The optical properties have been calculated within the random-phase approximation (RPA) using dipole matrix elements.
\\The NCs have been generated starting from a betacristobalite-SiO$_2$ matrix, by removing all the oxygen atoms inside a sphere whose radius determines the NC size. The so-obtained Si-NC, embedded in the SiO$_2$, presents perfectly coordinated atoms and the same Si-Si distance of betacristobalite, corresponding to about 3.2\,\AA. The NCs have been de-embedded by removing all the external Si/O atoms, while keeping the first shell of O atoms forming the interface. Finally, each interface oxygen has been passivated by an hydrogen atom. The resulting Si-NCs are OH-terminated, with a number of OH groups variable and dependent on the size of the spherical cutoff. In particular, the ratio of the number of oxygen by the number of Si atoms to which they are connected can vary from 1 to 3, and is named here the \textit{oxidation degree}, $\Omega$.\cite{PRB2} After the ionic relaxation the Si-Si bond length approached the bulk value of 2.34\,\AA, while the Si-O-H bonds formed angles of about 115\,$\degree$. 
Examples of so-obtained NCs are shown in Fig.~\ref{fig.structures}a. In addition to the OH-terminated NCs we have produced a set of H-terminated NCs by replacing the OH groups with hydrogens and then re-relaxing the structures (Fig.~\ref{fig.structures}b). Following the procedure outlined above we have generated the following set of NCs:
Si$_{17}$(OH)$_{36}$, Si$_{26}$(OH)$_{48}$, Si$_{29}$(OH)$_{36}$, Si$_{32}$(OH)$_{56}$, Si$_{35}$(OH)$_{36}$, Si$_{47}$(OH)$_{60}$, Si$_{61}$(OH)$_{66}$, Si$_{71}$(OH)$_{108}$, Si$_{87}$(OH)$_{76}$, Si$_{109}$(OH)$_{108}$, Si$_{147}$(OH)$_{100}$. The H-terminated counterparts of the above NCs set have been also generated, with the additional Si$_{293}$H$_{172}$.
\begin{figure}[b!]
  \centering
  \includegraphics[draft=false,height=\columnwidth,angle=270]{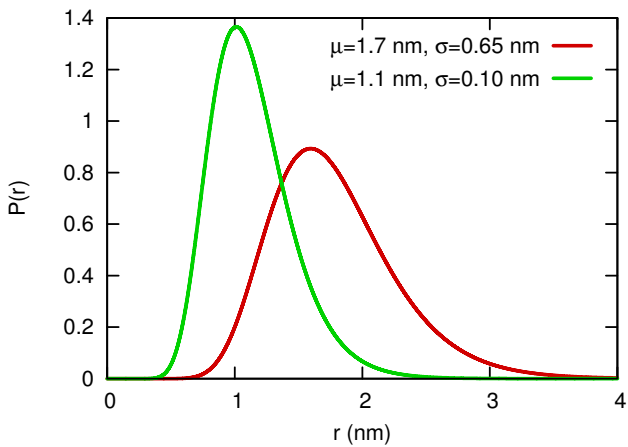}
  \caption{\small NC-radius distributions expressed by Eq.~\ref{eq.logdistrib}, with parameters obtained by fitting on experimental data of Ref.~\onlinecite{vinciguerra} for two different samples. The legend reports the curve parameters. }\label{fig.distributions}
\end{figure}

In common experiments the samples are characterized by a log-normal distribution in the NCs radius $r$ expressed by:\cite{lockwood}
\begin{equation}\label{eq.logdistrib}
 P(r)=\frac{1}{Sr\sqrt{2\pi}}\exp\left[\frac{-\left(\,\ln(r)-M\right)^2}{2\,S^2}\right]~,
\end{equation}
where the mean and the variance of the distribution can be related to $M$ and $S$ by $\mu$\,=\,$\exp(M+S^2/2)$ and $\sigma$\,=\,$\exp(S^2+2M)[\exp(S^2)-1]$, respectively.
\\Two realistic NC-radius distributions with mean radiuses of 1.7\,nm and 1.1\,nm obtained from fitting the experimental data of Ref.~\onlinecite{vinciguerra} using Eq.\,\ref{eq.logdistrib} are reported in Fig.~\ref{fig.distributions}. Clearly, by varying the silicon layer thickness and the annealing conditions it is possible to obtain samples with distributions of different mean and variance. In the case of freestanding NCs it is possible to obtain narrower distributions by using filters that determine the Si-NC diameter.\cite{mangolini,lockwood-meldrum,hannah}

\noindent When not specified, we assume eV and nm the default units of energy and distance, respectively.

\section{Absorption}\label{sec.absorption}
\noindent In this Section we aim at describing the absorption spectrum (here represented by the imaginary part of the dielectric function) of the ensemble by summing the individual NC spectra with weights given by Eq.\,\ref{eq.logdistrib}.
\\First, the DFT-RPA complex dielectric function of the NCs, $\varepsilon$, have been calculated for all the relaxed structures. In doing that, we have omitted the vacuum states, i.e. the conduction states of energy equal or above the vacuum energy $E_{vac}$. An estimate of $E_{vac}$ can be calculated by properly aligning the eigenvalues after applying the Makov-Payne correction to the total energy.\cite{makov-payne} In alternative, the vacuum states are identifyable by an inverse-participation-ratio (IPR)\cite{APL} value well-below a certain threshold. For each system, we have evaluated $\varepsilon$ removing the vacuum states by a crosscheck of both the above methods.
\\Second, we must consider that the calculated absorption is normalized over the volume of the simulation cell. Therefore, by means of effective-medium approximation (EMA) we have retrieved the NC-related dielectric function $\varepsilon_{nc}$ from the calculated one:
\begin{equation}\label{eq.cellvacuum}
 \varepsilon_{nc}=\frac{\varepsilon -(1-f)}{f}~,
\end{equation}
where the so-called \textit{filling factor}, $f$, is given by the ratio of the NC volume and the simulation cell volume
\begin{equation}\label{eq.filling-factor}
f = V_{NC}/V_{cell} ~.
\end{equation}
Given the spherical shape of the NC (by construction), we may simply relate the NC volume to its diameter $d$ by:
\begin{equation}\label{eq.V_NC(d)}
 V_{NC} = \frac{4\pi}{3} \left( \frac{d}{2} \right)^3~.
\end{equation} 
The problem of correctly define $d$ is not trivial, especially for small NCs, and requires some discussion. In a previous work\cite{PSSB} we have demonstrated that the dielectric function of the Si/SiO$_2$ embedded system $\varepsilon_{tot}$ is separable through the EMA into $\varepsilon_{nc}$ and the dielectric function of the hosting dielectric matrix $\varepsilon_{h}$. From such calculation we revealed that the true inclusion is not formed by the sole NC but by the NC+interface system. Similarly to Eq.~\ref{eq.cellvacuum}, the separation of the dielectric function was possible provided that a correct filling factor was introduced into the calculation. Therefore, by using the EMA it is possible to calculate an ``effective'' $d$ for a set of embedded NCs. Following this picture we have verified that the relationship
\begin{equation}\label{eq.d(rho)}
 d^* = 2\cdot\left( \frac{N_a}{\rho\frac{4}{3}\pi} \right)^{\frac{1}{3}}
\end{equation} 
returns the correct filling factor when $\rho$ corresponds to the atomic density of bulk Silicon, $\rho$\,=\,$8/(5.43\,\text{\AA})^3$\,$\simeq$\,$0.05\,\text{\AA}^{-3}$, and $N_a$ corresponds to the total number of atoms in the NC (Si+O+H).
\\Thus $\varepsilon_{nc}$ for each NC of the set has been calculated using Eq.~\ref{eq.cellvacuum} with the respective filling factor derived from Eqs.~\ref{eq.filling-factor}-\ref{eq.d(rho)}.
\\Note that the diameter resulting from Eq.~\ref{eq.d(rho)} is larger than the maximum distance between two atoms belonging to the same NC, widely used in literature to represent the NC diameter. Therefore, for the sake of comparison with other works, we distinguish the ``optical'' $d^*$ defined to to correctly evaluate $\varepsilon_{nc}$, by the ``physical'' $d$ that characterize the NC size, entering Eq.~\ref{eq.logdistrib}.

\begin{figure}[t!]
  \centering
  \includegraphics[draft=false,height=\columnwidth,angle=270]{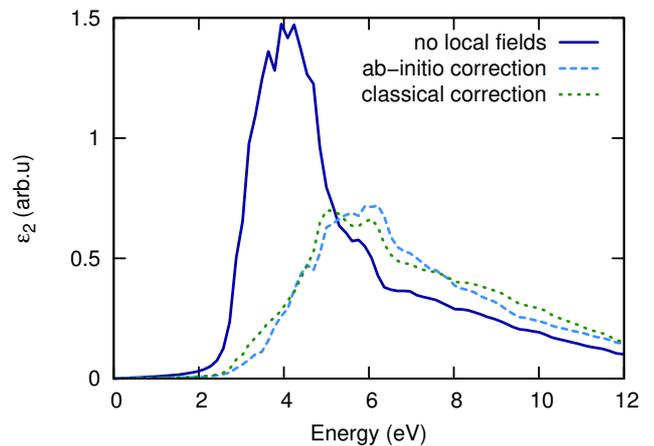}
  \caption{\small Imaginary part of dielectric function of  Si$_{32}$(OH)$_{56}$ NC without LF (solid curve), with LF correction calculated by ab-initio (dashed curve), and with LF correction calculated by Eq.~\ref{eq.LF} (dotted curve). }\label{fig.LF}
\end{figure}

\begin{figure}[t!]
  \centering
  \includegraphics[draft=false,width=\columnwidth,angle=0]{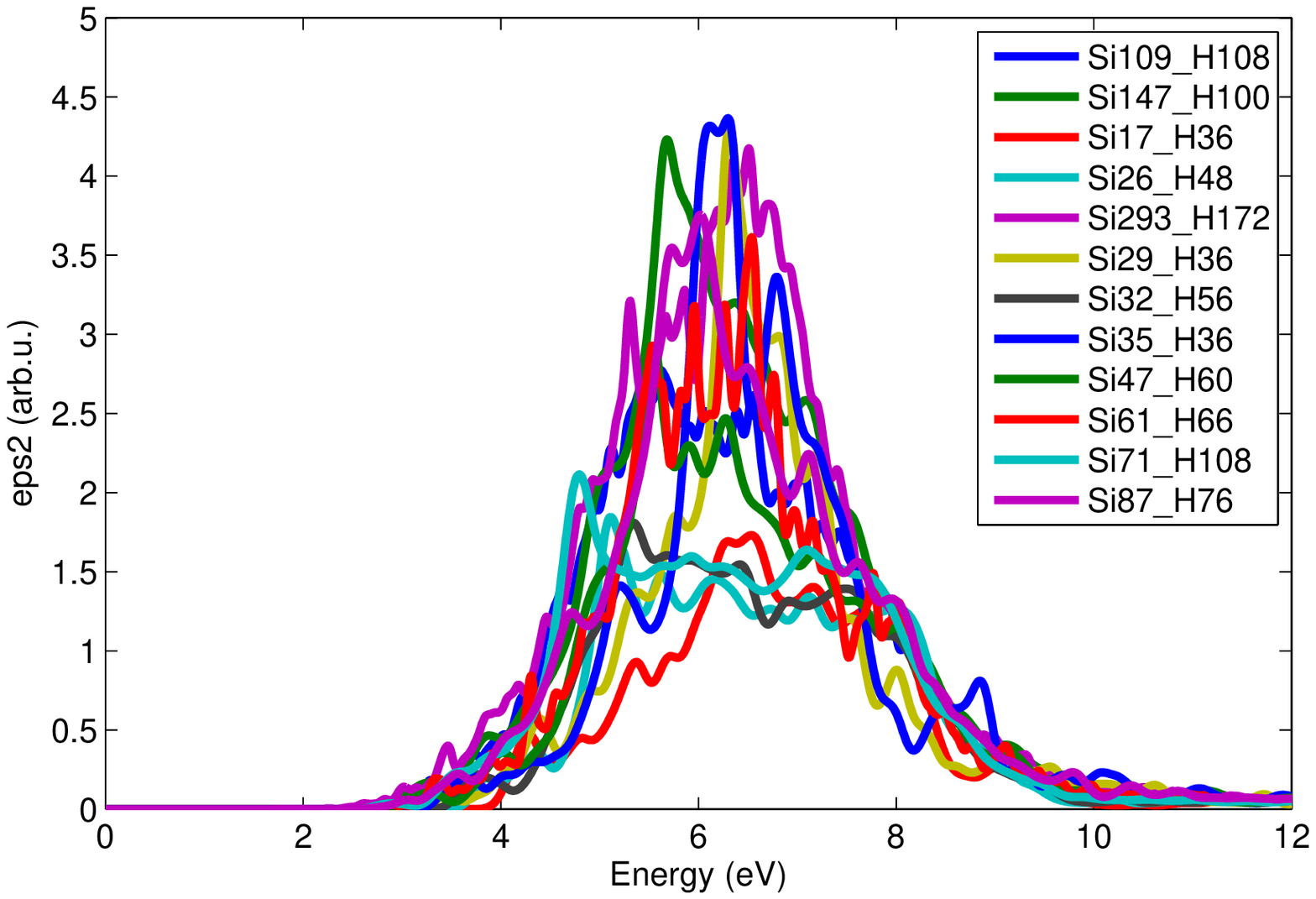}\\
  \includegraphics[draft=false,width=\columnwidth,angle=0]{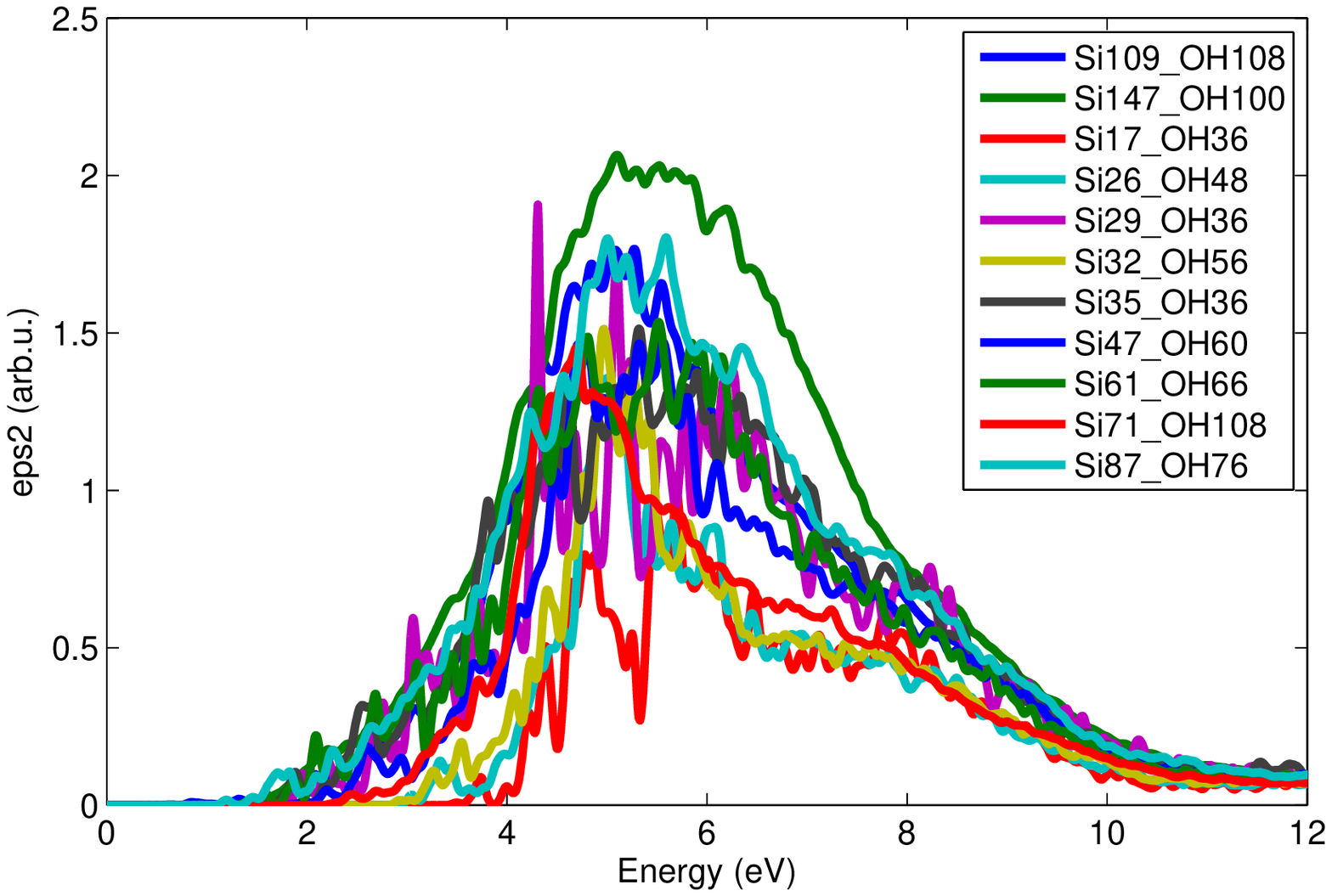}
  \caption{\small Absorption spectra, calculated using Eq.~\ref{eq.LF}, of the H (top panel) and OH (bottom panel) NCs set. }\label{fig.abs-all}
\end{figure}

Another crucial aspect of optical absorption in nanostructures concerns the effect of local fields (LF). While previous many-body calculations on Si-NCs have shown that self-energy corrections and electron-hole Coulombic corrections nearly cancel out each other\cite{PRB1} yielding fundamental gaps and absorption spectra close to the independent-particle calculated ones, the effect of LF has been reported to severely modify the absorption profile, as for embedded like as for freestanding NCs of every size.\cite{PRB4} Unfortunately, the ab-initio calculation of the full dielectric response requires a computational effort that increases dramatically with the system size, setting a strong limit on the maximum processable NC $d$.
\\To circumvent the above limitation we make use of the EMA in order to easily include the LF correction to the dielectric function. Since in NCs the LF are mostly given by surface polarization effects,\cite{PRB4} we make use of the \textit{Clausius-Mossotti} equation in order to describe the polarizability $\alpha$ of a dielectric sphere with dielectric constant $\varepsilon$ and volume $V$, embedded into a background with dielectric constant $\varepsilon_0$:
\begin{equation}\label{eq.clausius-mossotti}
 \alpha = 3\,V\,\varepsilon_0\,(\varepsilon-\varepsilon_0)/[4\pi(\varepsilon+2\varepsilon_0)]~.
\end{equation} 
Then, the LF-corrected $\varepsilon$ is given by
\begin{equation}\label{eq.LF_generic}
 \varepsilon_{LF} = \varepsilon_0 + \frac{4\pi\alpha}{V} = \varepsilon_0\frac{4\,\varepsilon_{nc}-\varepsilon_0}{\varepsilon_{nc}+2\,\varepsilon_0}~.
\end{equation}
Finally, since our NCs are embedded in vacuum, by posing $\varepsilon_0$\,=\,1 we get
\begin{equation}\label{eq.LF}
 \varepsilon_{LF}=\frac{4\,\varepsilon_{nc}-1}{\varepsilon_{nc}+2}
\end{equation}
The validation of the Eq.~\ref{eq.LF} is performed by a comparison with $\varepsilon$ calculated by full-response ab-initio techniques, as described in Ref.~\onlinecite{PRB4}. In Fig.~\ref{fig.LF} such comparison is reported for the Si$_{32}$(OH)$_{56}$ case. We observe that Eq.~\ref{eq.LF} is able to produce a LF correction nicely matching the ``exact'' one, confirming the idea that LF are mainly due to classical polarization effects.\cite{ReiningLF}
\\Therefore, by applying Eq.~\ref{eq.LF} on the $\varepsilon_{nc}$ of all the NC set we have obtained the final corrected spectra whose (weighted) superposition shall describe the absorption of the ensemble (see Fig.~\ref{fig.abs-all}).
\\Since for our OH-terminated NC set we obtain 1.26\,nm\,$\leq$\,$d$\,$\leq$\,2.11\,nm while for the H-terminated NCs we have 0.87\,nm\,$\leq$\,$d$\,$\leq$\,2.24\,nm, in order to describe the ensemble we have set $\mu$\,=\,0.84\,nm and $\sigma$\,=\,0.01\,nm.
\\Then, given $N$ NCs of increasing $d$ from $d_1$ to $d_N$, the weight $W_i$ for the $i$-th NC is expressed by
\begin{equation}\label{eq.weights}
	W_i=\int_{(d_i+d_{i-1})/2}^{(d_i+d_{i+1})/2}P(r)\,dr~,
\end{equation}
in which $d_0$\,=\,-$d_1$ and $d_{N+1}$\,=\,$\infty$.
\begin{figure}[t!]
  \centering
  \includegraphics[draft=false,height=\columnwidth,angle=270]{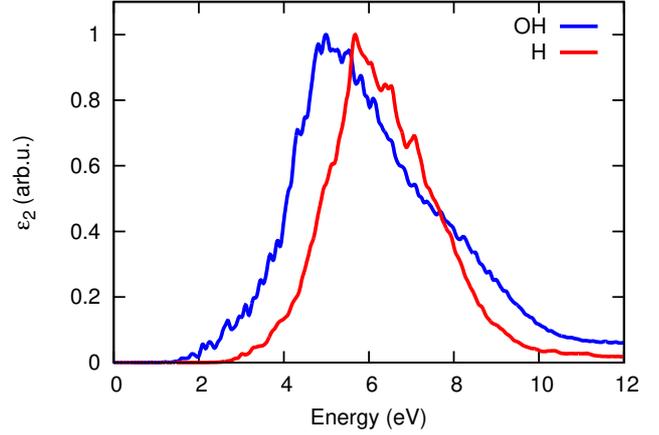}
  \caption{\small Calculated absorption spectrum of the NC ensemble made by H-terminated (red curve) and OH-terminated (blue curve) with radius distribution parametrized by $\mu$\,=\,0.84\,nm and $\sigma$\,=\,0.01\,nm. }\label{fig.abs-ensemble}
\end{figure}
\\Finally, the absorption of the ensemble is simply given by the weighted sum of the individual absorptions, through
\begin{equation}
	\varepsilon_{tot}=\sum_{i=1}^N \,W_i\,\varepsilon^i_{LF}~~.
\end{equation} 
The so-obtained spectra for the H- and OH-terminated NC ensembles are reported in Fig.~\ref{fig.abs-ensemble}. The calculated absorption is characterized by a triangular-like main peak centered at about 5.7\,eV for the H-terminated NCs, and at about 5.0\,eV for the OH-terminated ones. The $\sim$0.7\,eV shift is due to the well-known difference in the confinement capability of H w.r.t.\ OH passivation.\cite{PRB2,conibeer}
\\The calculated spectrum profile of OH-terminated NCs is in general agreement with experimental observations on NCs embedded in SiO$_2$ with average $d$ of 1\,nm\,\cite{gallas}, apart of a red-shift in the maximum of the experimental spectra that should be associated to the strain induced by the embedding matrix on the NCs.\cite{PRB2}

\section{Emission}\label{sec.emission}
In this Section we aim at evaluating the optical emission spectrum of the NC ensemble as sum of the emissions of the individual NCs.
\\Recently, some studies on the PL spectra of individual Si NCs have reported linewidths of 2\,meV at T\,$\simeq$\,35 K, clearly below $k_BT$ at this temperature, demonstrating true quantum dot PL emission characteristics.\cite{sychugov,empedocles}
\\This very important step permits us to modellize the PL of single NCs using an atomic-like response. Following the above assumption we have described the emission of a single NC through a gaussian profile,
\begin{equation}\label{eq.gaussian}
\begin{aligned}
 G(E) &= I(E_g)/(\gamma\sqrt{2\pi})\times \\
  &\qquad\times\exp\left[-\frac{1}{2} \,\left(\frac{E-E_g}{\gamma}\right)^2\right]~,
\end{aligned}
\end{equation}
in which the intensity $I$ and the position $E_g$ of the emission peak can be associated to the radiative recombination rate (RR) of the NC and to its energy gap, respectively. The linewidth of the peak, $\gamma$, is fixed to 150\,meV following experimental observations.\cite{sychugov,valenta}

Differently than the previous Section, by expressing $I$ and $E_g$ as a function of the NC size, in this case we can perform the summation using any kind of NC distribution. The assumption is that the analitic expressions of $I(d)$ and $E_g(d)$ are applicable also outside the fitting range.

\begin{figure}[t!]
  \centering
  \includegraphics[draft=false,height=\columnwidth,angle=270]{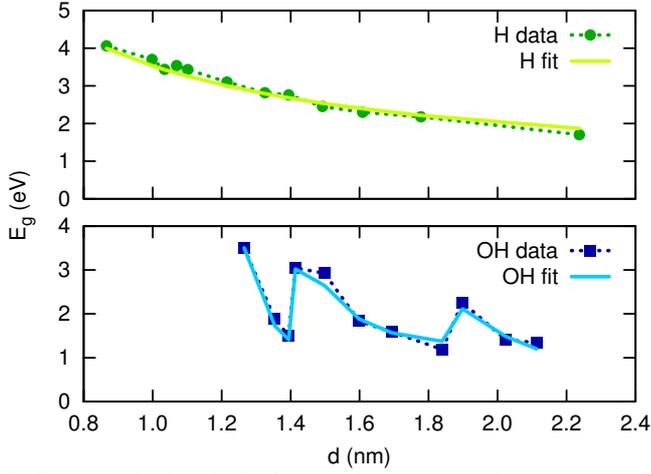}
  \caption{\small Calculated $E_g$ for the hydrogenated (upper panel) and OH-terminated (lower panel) NCs. Solid lines report the fit of the data, expressed by Eqs.~\ref{eq.E(d)}-\ref{eq.E(d,Omega)}. }\label{fig.gapformula}
\end{figure}

In the simplest picture, the NC energy gap $E_g$ is determined by the diameter $d$ following the quantum confinement (QC) picture:
\begin{equation}\label{eq.QC}
E_g=1.1+\alpha\,d^{-\beta}~~.
\end{equation} 
Eq.~\ref{eq.QC} corresponds to the particle-in-the-box model in which $\beta$ assumes the maximum value of 2 in the case of an infinite potential barrier.\cite{lockwood} However, in the presence of strongly polar interface terminations (e.g. OH) the interface has a major impact on the electronic structure of Si NCs.\cite{PRB2,conibeer} In this case the above model no longer works and one must take into account the specific configuration of the interface in order to correctly estimate $E_g$. In Ref.~\onlinecite{PRB2} we demonstrated the presence of a strong correlation between $E_g$ and $\Omega$ in the case of OH-terminated NCs. As a consequence, for such systems we must derive a $E_g(d,\Omega)$ relationship, while for H-terminated NCs a simpler $E_g(d)$ can be employed.
\\Therefore, we have calculated the $E_g$ of all the systems and fitted the data as a function of the NC size (see Fig.~\ref{fig.gapformula}). For the H-terminated NCs we have obtained
\begin{equation}\label{eq.E(d)} 
 E_g(d) = 0.53 + 3.0/d~,
\end{equation} 
while for the OH-terminated NCs we have derived the expression
\begin{equation}\label{eq.E(d,Omega)} 
\begin{split}
 E_g(d,\Omega) = 0.53 + 1.5/d + 1.8\,(\Omega-1.41)/d^2
\end{split}
\end{equation}
in which the last term produces the correction due to the variation of $\Omega$. We note that such term vanishes out rapidly at large $d$, suggesting that this effect is limited to small NCs. In agreement with other works,\cite{kim,vasiliev,zhou} this term becomes negligible at $d$ $\simeq$3\,nm, a threshold above which the gap can be considered purely QC driven.
\\Eqs.~\ref{eq.E(d)}-\ref{eq.E(d,Omega)} tend to 0.53\,eV for asymptotically large NCs, being the DFT-LDA energy gap of bulk-silicon. In principle, one may apply a so-called \textit{scissor operator} of 0.6\,eV in order to force the calculated bulk value to the experimental one. Besides, since in our case we work very far from the bulk limit (i.e. small NCs) no such correction to the computed gap have been applied at this stage.
\\The $d^{-1}$ trend of the gap has been chosen in agreement with other calculations,\cite{delley} while the reason for the value of 1.41 in Eq.~\ref{eq.E(d,Omega)} will be clarified in the following. All the remaining parameters resulted from data fitting.
\\By looking at Fig.~\ref{fig.gapformula} it is possible to compare the computed $E_g$ with the ones given by the fit. Clearly, Eqs.~\ref{eq.E(d)}-\ref{eq.E(d,Omega)} are able to estimate with good precision the $E_g$ value of each NC.

\begin{figure}[t!]
  \centering
  \includegraphics[draft=false,height=\columnwidth,angle=270]{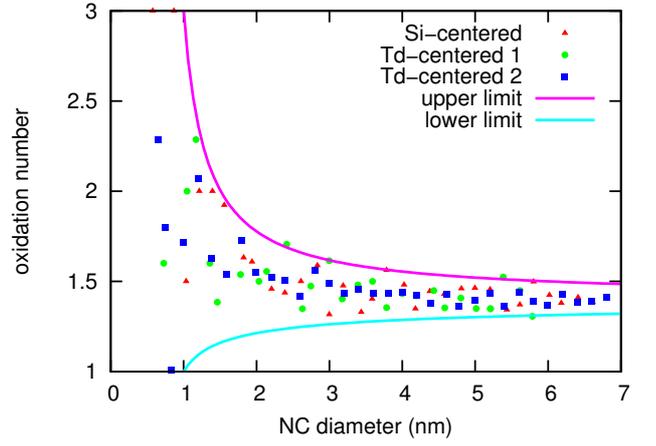}
  \caption{\small $\Omega$ values for a set of NCs generated with random $d$ and centered on a Si-atom or on two different $T_d$-interstitials. Upper and lower curves define the maximum and minimum achievable $\Omega$ at each $d$. }\label{fig.omegas}
\end{figure}

Next, we must determine the $\Omega$ value to be used in Eq.~\ref{eq.E(d,Omega)}. To do that we have generated a large number of NCs using cutoff-spheres of different $d$, and with different centering sites: on a Si-atom or on two different $T_d$-interstitials.
In this way we have calculated the $\Omega$ value of a large number of NCs with $d$\,$\leq$\,7\,nm, as reported in Fig.~\ref{fig.omegas}. Clearly, at large $d$ the oxidation degree tends to a well defined value of about 1.41\,$\simeq$\,$\sqrt2$ 
(used in Eq.~\ref{eq.E(d,Omega)}), while at very small $d$ it tends to increase on the average, in agreement with experimental observations.\cite{kim}
A fit of the maximum-possible and minimum-possible $\Omega(d)$ yields
\begin{gather}\label{eq.omegas}
 \left.\begin{aligned}
 \Omega_{min}(d) &= 1.41-0.225/\sqrt{d-0.7} \\
 \Omega_{max}(d) &= 1.41+0.477/(d-0.7)
 \end{aligned}
 ~\right\}~~,~~\text{d$\geq$1\,nm} \nonumber \\
 \left.\begin{aligned}
 \Omega_{min}(d) &= 1 \\
 \Omega_{max}(d) &= 3
 \end{aligned}
 ~\right\}~~,~~\text{d$<$1\,nm}
\end{gather}
As depicted by solid curves of Fig.~\ref{fig.omegas}, $\Omega_{min}(d)$ and $\Omega_{max}(d)$ limit the possible value of $\Omega$ that a randomly generated NC of given $d$ can assume.

Next, to determine the amplitude of the emission peak, we make use of the RR calculated in Ref.~\onlinecite{PRB3}. For the H-terminated NCs we get
\begin{eqnarray}\label{eq.I(E)_H}
 I(E) &= 7.7\cdot10^6\,(E-2.47)^{1.19}~~,~~ E>2.47\,\text{eV} 
\end{eqnarray} 
while for the OH-terminated ones
\begin{eqnarray}\label{eq.I(E)_OH}
 I(E) &= 3.7\cdot10^5\,(E-0.71)^{2.94}~~,~~ E>0.71\,\text{eV} 
\end{eqnarray}

\begin{figure}[t!]
  \centering
  \includegraphics[draft=false,height=\columnwidth,angle=270]{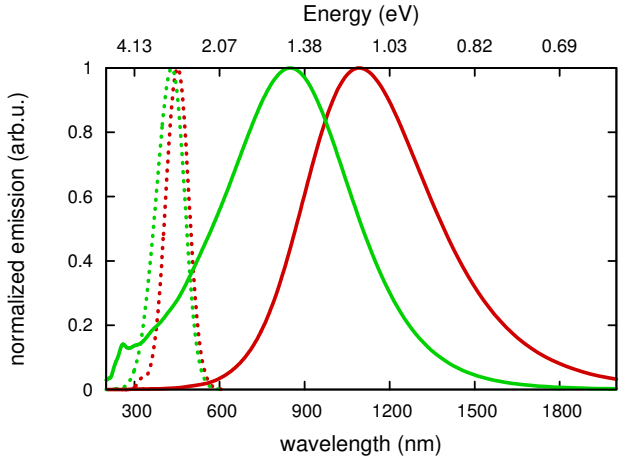}
  \caption{\small Optical emission spectra of NC ensembles with radius distributions parametrized by $\mu$\,=\,1.7\,nm,\,$\sigma$\,=\,0.65\,nm (red curves) and $\mu$\,=\,1.1\,nm,\,$\sigma$\,=\,0.10\,nm (green curves). Solid (dashed) curves represent spectra of OH-terminated (H-terminated) NC ensembles. }\label{fig.emissions_nm}
\end{figure}

We have finally generated a large set of NCs ($N$\,=\,10$^6$), with $d$ distributed using $P(x)$ with realistic parameters (see Fig.~\ref{fig.distributions}) and randomly-generated $\Omega$\,=\,[$\Omega_{min}$..$\Omega_{max}$]. For each NC, the $G(E)$ has been calculated using $E_g$ from Eqs.~\ref{eq.E(d)}-\ref{eq.E(d,Omega)}, and $I(E_g)$ from Eqs.~\ref{eq.I(E)_H}-\ref{eq.I(E)_OH}. Finally, the sum over the $N$ $G(E)$ has been performed, forming the emission spectrum (i.e. PL) of the ensemble, as reported in Fig.~\ref{fig.emissions_nm}.
\\The emission spectra of the OH-terminated NC ensembles are characterized by broad profiles peaked at about 850\,nm (1.45\,eV) and 1090\,nm (1.15\,eV), to be compared with the measured 800\,nm (1.55\,eV) and 900\,nm (1.38\,eV) of experimental samples with corresponding distributions.\cite{vinciguerra}
\\The small peak at about 200\,nm {present in the emission (green curve) relative to the OH terminated ensemble} is clearly related to the wide variation of $\Omega$ in small NCs, being not present in the sample with larger NCs.
\\The calculated spectra are slightly red-shifted w.r.t.\ the measured ones, with the largest red-shift appearing on the ensemble containing larger NCs, suggesting that the used equations may loose accuracy when applied far from the fitting range. In particular, while in the strong QC regime the gap value of bulk-Si in Eq.~\ref{eq.E(d,Omega)} plays a minor role in determining $E_g$ (since also other mechanisms concur, e.g. electron-hole interaction),\cite{PRB1} at large $d$ it becomes a critical parameter, and its underestimation generates as a red-shift that increases with $\mu$.
\\The latter effect is also expected to contribute to the excessive linewidth affecting the spectra. Besides, the sharpness of the peaks of H-terminated NC ensembles indicates a connection between the broadening and the dependence of $E_g$ with $\Omega$ in OH-terminated NCs. In fact, since the emission intensity increases at smaller $d$, the large variation of $E_g(\Omega)$ at small $d$ produces NCs strongly emitting in a wide energy range. This is not the case of hydrogenated NCs, presenting $E_g$ with very small dispersion in energy at low $d$, and a correspondingly narrow PL peak.
\\A second possible motivation of the oversized linewidths could arise from the NC-NC interaction mechanisms, completely neglected in the present work. For example, we expect that the inclusion of F\"orster and tunneling ``energy migration'' interactions in the model would consistently reduce the PL linewidths.\cite{lockwood}

\begin{figure}[t!]
  \centering
  \includegraphics[draft=false,height=\columnwidth,angle=270]{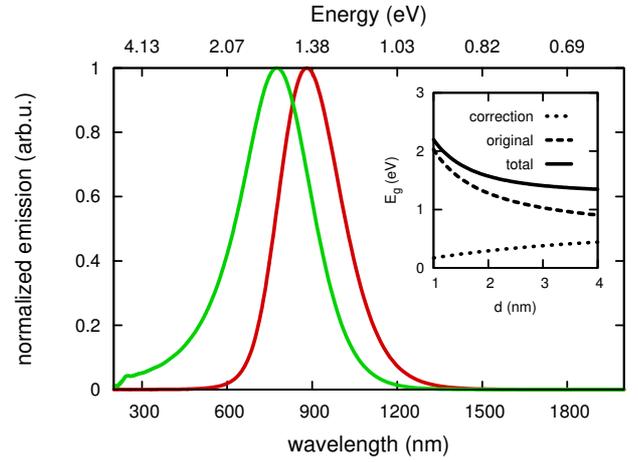}
  \caption{\small Optical emission spectra of OH-terminated NC ensembles with radius distributions parametrized by $\mu$\,=\,1.7\,nm,\,$\sigma$\,=\,0.65\,nm (red curves) and $\mu$\,=\,1.1\,nm,\,$\sigma$\,=\,0.10\,nm (green curves) obtained using $E_g$ corrected through Eq.~\ref{eq.scissor}.  The inset reports $E_g(d,1.41)$ (dashed curve), $C(d)$ (dotted curve), and their sum (solid curve). }\label{fig.emissions_nm_scissor}
\end{figure}

To better understand the role of the mechanisms in play, we have introduced an additional corrective term to Eq.~\ref{eq.E(d,Omega)} in order to reproduce the experimental bulk-Si value at large $d$. The proposed corrective function is
\begin{equation}\label{eq.scissor}
	C(d) = 0.6\,\left[1-e^{-d/3}\right]~,
\end{equation}
in which the exponential argument has been chosen so that $E_g(d,\Omega)$+$C(d)$ returns a trend of the gap comparable to the experimental one\cite{conibeer} when $\Omega$ is set to its asymptotic value.
\\The inset of Fig.~\ref{fig.emissions_nm_scissor} reports $E_g(d,1.41)$, $C(d)$, and their sum, while the main Figure reports the spectra of the OH-terminated NC ensembles corrected using Eq.~\ref{eq.scissor}.

The corrected spectra show much reduced linewidths, and peaks with maximum at 780\,nm (1.59\,eV) and 880\,nm (1.41\,eV), very close to the experimental counterparts. This indicates that NC-NC interaction mechanisms play a secondary role in the samples of Ref.~\onlinecite{vinciguerra}.
\\Besides, the comparison with PL of hydrogenated NCs supports the idea of addressing the broadness of the PL peak observed in experiments to the sensitivity of $E_g$ to the oxidation level at the interface, in particular for small NCs. The latter aspect emerges by comparing the present model with that of Ref.~\onlinecite{meier}, parametrized by experimental data. While at large $d$ the latter model show a nice agreement with the emission profiles of corresponding NC samples, it cannot reproduce the PL of the sample with the smallest NCs, presenting a very broad peak with much increased emission at high energy. In our opinion such feature is related to the large variation of $\Omega$ at small $d$, and to the fact that the smallest NCs present the highest radiative rates,\cite{meier} as expressed by Eq.~\ref{eq.I(E)_OH}. Therefore, a comprehensive model for optical emission should comprise the oxidation characteristics of the sample, as suggested by early experimental works.\cite{kanemitsu}

\begin{figure}[t!]
  \centering
  \includegraphics[draft=false,height=\columnwidth,angle=270]{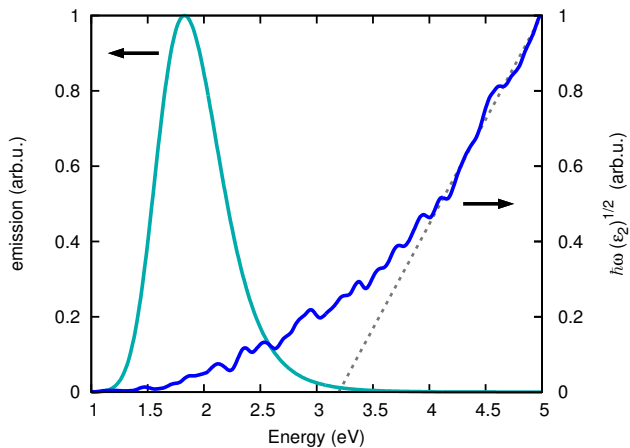}
  \caption{\small Optical emission (left curve) and absorption (right curve) spectra of OH-terminated NC ensemble with radius distribution parametrized by $\mu$\,=\,0.84\,nm and $\sigma$\,=\,0.01\,nm. The absorption curve is plotted using the Tauc method,\cite{Tauc} and the extrapolation of the Tauc gap is reported by the dotted curve. Both spectra have been corrected by including the $C(d)$ term of Eq.~\ref{eq.scissor}. }\label{fig.stokes}
\end{figure}
At last we compare, in Fig.~\ref{fig.stokes}, the absorption and emission spectra of the OH-terminated sample with radius distribution parametrized by $\mu$\,=\,0.84\,nm and $\sigma$\,=\,0.01\,nm. In the Figure, both absorption and emission have been calculated by including the $C(d)$ correction term. The absorption spectrum has been plotted following Tauc,\cite{Tauc} and a Tauc gap $E_T$\,=\,3.19\,eV has been obtained by a linear fitting ($E_T$\,=\,2.96\,eV for the uncorrected spectrum of the OH-terminated sample of Fig.~\ref{fig.abs-ensemble}). The maximum of the emission peak is positioned at 1.82\,eV.
\\Despite the several assumptions, the depicted results are in good agreement with the experimental outcomes showing that a very weak absorption exists in the region where luminescence peaks.\cite{nature_pavesi,gardelis,pavesimil,sgrigna} The origin of the large Stokes shift between absorption and emission peaks has been subject of intense debate from twenty years to date. While contributions from tunneling between NCs\cite{lockwood} and from structural deformation of the excited NCs\cite{leoprb,galliJACS,degoliprb} have been proposed, the most acknowledged contribution to the Stokes shift comes from associating emission and absorption to interface states and to quantum-confined states in the NC, respectively.\cite{nature_pavesi,leoprb,daldosso} Within the latter picture, the Tauc gap helps in distinguishing the absorption due to interface (surface) states ($E$\,$<$\,$E_T$) and due to NC states ($E$\,$>$\,$E_T$).
Evidently, since in our model the atomic-like emissions of the individual NCs are centered at $E_g$, the emission peak occurs entirely at $E$\,$<$\,$E_T$. This result is consistent with that of Ref.~\onlinecite{gardelis}, in which they report a Tauc gap of about 2.5\,eV and a PL peak centered at about 1.7\,eV for experimental samples made by SiO$_2$-embedded NCs with average $d$ smaller than 2\,nm. As already discussed in Sec.~\ref{sec.absorption}, the difference between the experimental and computed $E_T$ should be addressed to the SiO$_2$-induced strain on the NCs.\cite{PRB2}

\section{Conclusions}\label{sec.conclusions}
First, by performing DFT calculations on a set of Si-NCs, either passivated by H or OH, we have demonstrated the possibility of simulating the optical absorption spectrum of an ensemble of NCs with a realistic distribution in the size. The calculated spectrum is validated by a comparison with the experimental absorption of a corresponding sample. 
\\Second, a purely-analitical model for the optical emission of NC ensembles has been parametrized by DFT calculations and has been successively validated using experimental samples made by NCs with average radiuses of 1.1\,nm and 1.7\,nm. The presented model takes into account the oxidation degree of the NC, that appears particularly important for correctly describing the emission of the small NCs in the ensemble. Also, an important role of the SiO$_2$-induced stress has been confirmed, especially on the absorption, while a marginal role of the NC-NC interaction is deduced by the comparison with experiments.

\vspace{0.5cm}
\noindent \small{{\bf ACKNOWLEDGEMENTS:} Computational resources were made available by CINECA-ISCRA parallel computing initiative. We acknowledge financial support from the European Community's Seventh Framework Programme (FP7/2007-2013) under Grant No. 245977.}

\end{document}